\documentclass[6pt]{article}
\usepackage{indentfirst}
\usepackage{amsmath}
\usepackage{graphicx}
\usepackage{hyperref}
\usepackage{amssymb}

\author{Stanislav Srednyak}
\title{Bound state equations in Riemannian geometry}
\begin{document}
\maketitle

\abstract{
We study formulations of bound state (Bethe-Salpeter) equations on arbitrary Riemannian manifolds. We obtain a hierarchy of  equations for multipartice wave functions. These equations, at each number of particles, depend on certain choices of combinatorial origin, which together with the metric, define the equations completely. 
}

\section{Introduction}

Bound states in quantum field theory are mysterious objects, from mathematical point of view. The calculation of the simplest 1-loop correction to the simplest bound state - hydrogen atom - requires the use of hypergeometric functions and not all steps in the calculation have been done analytically ~\cite{Pachucki}. Calculations at 2 loop level are substantially more complicated ~\cite{Eides_book}, are scattered in many publications ( see  ~\cite{Eides1, Pachucki2, Yerokhin1}, to start with) and are not easy to follow for an average worker in the area of quantum field theory. 

Calculations in the theory of helim excited states are accordingly more complicated. One source of complexity stems from the fact that the mere formulation of quantization condition for non-perturbed wave functions involves Stokes phenomenon ~\cite{Lazutkin,Faddeev_Merkuriev}, which complete mathematical treatment is still lacking ( see ~\cite{Eremenko} for much easier anharmonc oscillator case, see ~\cite{Mochizuki,Kedlaya} for mathematical development regarding high dimensional Stockes phenomenon). It is known that classical 3-body systems are chaotic ~\cite{Lazutkin}. It has not been clarified what is the quantum counterpart of this (see however ~\cite{Turbiner}). Radiative corrections to helium bound states were considered in a number of works, see e.g. ~\cite{Pachucki_helium}. 

Bound state problem in QCD is a classic unsolved problem of our time. There exist various approximate approaches ( see e.g. ~\cite{review} ). Lattice approaches are very popular ~\cite{lattice}.

The remarkable feature of these developments is the absence of clear mathematical foundations of the theory. In the classic textbooks of the field ~\cite{Itzykson,Landau,Kinoshita} it is claimed that so called Bethe-Salpeter ~\cite{Bethe_Salpeter} equations are adequate to solve the problem completely. However, after the examination of the literature cited above it becomes clear that these calculations are in fact not based on the Bethe-Salpeter equations, but are rather approximate perturbative schemes, built around the notion of unperturbed wave function. One promising way to organize the perturbative series is provided by NRQED ~\cite{Caswell}.

On the mathematical level, it is not clear what is the meaning of the Bethe-Salpeter equations in therms of the structure of the underlying space time. The author is not aware of attempts to formulate such equations on Riemannian manifolds. One would expect that there is a hierarchy of wave functions of increasing number of variables that denote points in the underlying manifold , and increasing spin. This is of course makes sense only if one is willing to assume that there is certain manifold, that underlies physical processes, and that it is possible to interpret particle processes in terms of probability disctributions on tuples of manifolds. This is not at all obvious, as it is imaginable ( and probably true) that true evolution of particle systems happens in certain infinite-dimensional, functional , space, and is only being projected to finite dimensions in the observed states. 

In this paper, we prefer to be modest and focus on the study of analogies of Bethe-Salpeter hierachy in Riemannian geometry. We make the observation that it is possible to define a hierarchy of wave functions on arbitrary Riemannian manifold. These multicomponent equations cane be coupled in an intricate way. There is considerable freedom in coupling these equations. This freedom can be encoded in a universal algebra quite similar to the homotopy algebras extensively studied in mathematics ~\cite{homotopy_alg}, but still quite different. This algebra stems from the structure similar to Feynman diagrams. 

The resulting equations form an intricate system of integro-differential equations, which suffers from the usual problem of ultraviolet divergences. The kernels of integration have singularities on various multidiagonals, and need to be regularized. We note that the problem of renormalization in Riemannian geometry is not completely solved in the literature (except for 3d manifolds ~\cite{Axelrod_Singer}).

\section{Classical Bethe-Salpeter equations on $M^4$}

In this section we expling the most classic version of Bethe-Salpeter equations, as formulated in ~\cite{Bethe_Salpeter}. It is assumed that the spacetime is the Minkowski space $M^4$. This liner structure is very essential for the whole formulation. The translation invariance dictates that the Green's functions of fields are functions of the difference between the two arguments. This is a very strong assumption. The equation involves two fields - spinor field $\psi_\alpha(x)$ and gauge field $A_\mu(x)$. The spinor index denotes the spinor in the representation of the double cover $SU(2)\times SU(2)$ of the 4-dimensional Lorenz group, so the spinor is a section of rank 4 bundle. We only consider $U(1)$ gauge fields ( the formulation for other gauge groups is a non-trivial problem, see however ~\cite{Roberts} for an approach based on Schwinger-Dyson hierarchy ). 

The basic objects in which the theory is formulated are causal Green's functions. The definition of these functions is somewhat complicated (see ~\cite{Bogoliubov}) and relies essentially on the Lorenz geometry. For our purposes we only note that these functions satisfy the following equations
\begin{gather*}
(i\hat{\partial}_x+m)_{\alpha,\gamma}S_{\gamma\beta}(x-y)=\delta_{\alpha\beta}\delta(x-y) \\
\Box G_{\mu\nu}(x-y)=\delta(x-y)
\end{gather*}
These equations are ambiguous on the non-compact Minkowski space and it is necessary to fix boundary conditions at temporal infinity. Two possible choices are possible - the ones that fixes the function to vanish in remote past, which correspond to advanced boundary conditions, and the one that corresponds to the vanishing in the forward light cone , that corresponds to retarded boundary conditions. We will denote these functions by superscripts $adv$ or $ret$ respectively. The causal Greens functions are then defined as
\begin{gather*}
G(x)=G^{ret}(x)\theta(t)-G^{adv}(x)\theta(x)
\end{gather*}

Starting from the functions $S(x) ,G(x)$ it is possible to define correlation functions, using Feynman diagrams ~\cite{Landau}. We will be espesially interested in 2electron $\rightarrow$ 2electron kernel, as it plays a prominent role in the formulation of the Bether-Salpeter equation for positronium - the basic system we are interested. This kernel has the following signature $K_{\alpha,\beta,\gamma,\zeta}(x,y,u,v)$. Due to translation invariance it depends only on 3 spacetime variables. It has 4 spinor indices. In the original approach, all perturbative contributions to this kernel we considered. These contributions are enumerated by Feynman diagrams with 4 external electron lines. A few contributions to $K$ are listed below
\begin{equation}
\int dzdu S(x,z)\gamma_{\mu} S(z,u)G_{\mu\nu}(u,v)S(y,w)\gamma_{\nu} S(w,u)
\end{equation}
This expression corresponds to the diagram with single photon line.
\begin{gather*}
\int d^4ad^4bd^4cd^4d S(x,a)\gamma_\mu S(a,c) \gamma_\nu S(c,u) G_{\mu\mu'}(a,b) \times\\
\times G_{\nu\nu'}(c,d) S(y,b)\gamma_{\mu'}S(b,d)\gamma_{\nu'}S(d,v)
\end{gather*}
which corresponds to the box diagram.

The Bethe-Salpeter equation is formulated in terms of 2-particle wave function $\psi_{\alpha,\beta}(x,y)$ of the pair electorn-positron. This function has two spinor indices and each of them corresponds to either electron or positron space. The equation is
\begin{equation}
(\hat{\partial_x}+m)(\hat{\partial_y}+m)\psi(x,y)=\int d^4ud^4v K(x,y,u,v)\psi(u,v)
\end{equation}
This is an integro-differential equation. 

Methods of solution invariably rely on the choice of factorized zero-order approximation for the function $\psi(x,y)$.
 This function ostensibly depends on two time variables. In the known approaches to solution, these two times are chosen to be equal. It is not known to the author how to avoid explicitly making this choice and still obtain the solution to the spectral problem (see however promising approaches ~\cite{multitime})

It is possible to imagine multiple bound states in QED, which contain both electrons, positrons and photons. There are good reasons to consider such mixed bound states. It is known that radiative corrections involve infrared divergences and these divergences do indeed plague higher order computations. It is necessary to introduce infra-red photons in the final state to cancel the loop contributions. This presicely corresponds to consideration of factorizable mixed electron-photon wave function. 

Here we provide a few examples of Bethe-Salpeter equations for mixed wave function $\psi_{\alpha,\beta\mu}(x,y,z)$
\begin{equation}
(i\hat{\partial_x}+m)(i\hat{\partial_y}+m)\Box_z\psi_{\alpha,\beta\mu}(x,y,z)=\int K(x,y,z,u,v,w)\psi(u,v,w)d^4u d^4v d^4w
\end{equation}
This equation involves a kernel that has 6 external lines - 4 electron, and 2 photon lines.

\section{Bether-Salpeter hierarchy in Riemannian geometry}

In this section, we demonstrate that to any Riemannian manifold there corresponds a hierarchy of functions and equations for them, that generalizes the Bethe-Salpeter equations discussed above and the classical equations for Green's functions and harmonic forms. This hierarchy and its solutions are therefore functionals of Riemannian metric and must have geometric meaning. 

Instead of working with spinors and vectors as we did before, we choose to formulate our equations for forms, to simplify our discussion. The basic functions for which the equations are defined are forms on tuples of our original manifold $M^d$
\begin{equation}
\omega_{I_1,...,I_n}(x_1,...,x_n),\ x_i \in M,\ I_i={\mu^{(i)}_1,...,\mu^{(i)}_{s_i}}
\end{equation}
These forms are analogies of multiparticle wave functions. 

These forms are elements of the space $\Omega^{I_1,...,I_n}(M\times...\times M)$. 

For these functions we can formulate eigenfunction equations of the form
\begin{equation}
(\Delta_1+...+\Delta_n)\omega^{(k)}(x_1,...,x_n)=\lambda_k \omega^{(k)}(x_1,...,x_n)
\end{equation}
index $k$ denotes the $k$-th eigenfunction ( we assume the spectrum is discrete. Simiral development can be carried out in the case of mixed spectrum).

For the field $\omega_k(x_1,...,x_n)$ we can obtain the corresponding Green's function 
\begin{gather*}
G_{I_1,...,I_n;J_1,...,J_n}(x_1,...,x_n;x'_1,...,x'_n)   =\\
\sum_k \frac{\omega_{I_1,...,I_n}^{(k)}(x_1,...,x_n) \omega_{J_1,...,J_n}^{(k)}(x'_1,...,x'_n)}{\lambda_k}
\end{gather*}

This picture corresponds to non-interacting theory. Now we turn to the introduction of interaction into this theory. We will take our motivation from Bethe-Salpeter theory and use integral kernels to add nonlinear interaction to the Laplace equations. Our equations will be of the form
\begin{equation}
(\Delta_1+...+\Delta_n )\omega_n(x_1,...,x_n)+K(x_1,...,x_n)[\omega_1,\omega_2,...]=\lambda \omega_n(x_1,...,x_n)
\end{equation}
Here $K(x_1,...,x_n)[\omega_1,\omega_2,...]$ is certain functional -interaction functional - of the multiparticle wave functions. Our goal is to define its structure.

To simplify the notation, we suppressed the tensor indices ( which can be restored from invariance condition). Using the set of solutions to this equation we define the Green's functions as follows
\begin{gather*}
G_{I_1,...,I_n;J_1,...,J_n}(x_1,...,x_n;y_1,...,y_n)=\\
\sum_k \frac{\omega^{(k)}_{I_1,...,I_n}(x_1,...,x_n)\omega^{(k)}_{J_1,...,J_n}(y_1,...,y_n)}{\lambda_k}
\end{gather*}

We now turn to the definition of the functional $K$. It is defined as the following formal expansion
\begin{gather*}
K(x_1,...,x_n)[\omega_1,\omega_2,....]=\\
=\sum_{k}\sum_{S_1,...,S_k} \int dy^{(1)}...dy^{(k)}K_{S_1,...,S_k}(x_1,...,x_n;y_{1,1},...,y_{1,m_1};....;y_{k,1},...,y_{k,m_k})\times\\
\times\omega_{S_1}(y_{1,1},...,y_{1,m_1})...\omega_{S_k}(y_{k,1},...,y_{k,m_k})
\end{gather*}
where the sum is performed over the number of variables and the tensorial structures that we wish to include in our theory. It is clear from this expression that there is considerable freedom in the formulation, as we are free to chose the signatures $S_i$.

The kernels $K_{S,S_1,...,S_k}(x_1,...,x_n;y_{1,1},...,y_{1,m_1};....;y_{k,1},...,y_{k,m_k})$ will be defined based on a generalization of Feynman diagrams. The conventional Feynman diagrams contain vartices of only one type, that is determined by the theory. Propagators are also fixed by the field content. We propose to generalize this construction and include effective propagators $G_{S,T}(x_1,...,x_n;y_1,...,y_n)$. These propagators carry the number of variables and the signature information. For each $n$ and each $S,T$ we have a propagator, and we define generalized Feynman lines by this data. The kernels $K_{S,S_1,...,S_k}(x_1,...,x_n;y_{1,1},...,y_{1,m_1};....;y_{k,1},...,y_{k,m_k})$ are sums over all modified Feynman diagrams, in which all possible dimensionalities of propagators, signatures of forms, and vertices are allowed.

Now we define the vertices of the theory. For each tuple $(\xi_1,S_1),...,(\xi_a,S_a)$ we will have a separate vertex, where $\xi_i=(x_{i,1},...,x_{i,q_i})$. We choose arbitrarily the splitting of the set $(x_{i,a})$ into $w$ groups $Y_1,...,Y_w$ of variables $y_b$ such that for each group $x_{i,a}=y_b$ within each group, and such that the corresponding parts of the tensor structures $S_a$ contract to form a volume form on the space $\{y_b\}=M$. As we see, there is considerable freedom in the definition of vertices. We will call such vertices as vertices of   $\{((x_{1,1},S_{1,1}),...,(x_{1,q_1},S_{1,q_1}));...;((x_{n,1},S_{n,1}),...,(x_{n,q_n},S_{n,q_n}))\} \rightarrow (y_1,...,y_w)$ type. It cooresponds to the interaction of the multiparticle effective fields $\omega_{R_1}(x_{1,1},...,x_{1,q_1}),...,\omega_{R_n}(x_{n,1},...,x_{n,q_n})$ Each such vertex corresponds to an integration of the variables $y_1,...,y_w$. One of the simplest type of the vertices is given by the splitting into triples of variables, which corresponds to the elementary QED vertex.

\section{The case of non-matching tensor structure at the vertex}

Our definition of interaction vertices can be extended to the case when the tensor structures for each of the copy of $M$ are not equal to the volume form, but equal to a tensor of some lower rank $r_s$. We still wish to keep integral representation for the vertices. We are therefore lead to consider integrations over submanifolds $V_{r_1,...,r_a}$ of appropriate dimensionality in the space $M\times...\times M$ of the vertex. The dimensionality of this manifold is determined by the ranks $r_i$ of the forms. We can consider this manifold to be immersion of a diffeomorphism type in our ambient space. 

We can formulate our modified Feynman rules as follows. The data for the theory is defined by basic field content $\omega_{\mu_1,...,\mu_s}(x)$ defined on $M$. From this data we constuct multi-field content, essentally as decribed in the previous section, obtaining forms on tuples of the manifold M, $\omega^{r_1,...,r_s}(x_1,...,x_s)$, where earch $r_i$ denotes tensor data $\mu^{(i)}_1,...,\mu^{(i)}_{r_i}$ that corresponds to the spatial variable $x_i$. The vertex is defined by the following data

1) Set of multifields that can interact in this vertex. We denote them as $\omega^{R_1}(x^1_1,...,x^1_{d_1}),...,\omega^{R_n}(x^n_1,...,x^n_{d_n})$. $R_i$ doneote the tensor structure $R_i=\{r^{(i)}_1,...,r^{(i)}_{d_i}\}$, where each $r^{(i)}_s$ is specified above.

2) Locality object. In the vertex we choose a set of variables $y_1,...,y_p$ and assignment $x^m_k \rightarrow y_l$ for each of the arguments $x^m_k$. This is our version of locality.

3) Using the assignment $x^m_k \rightarrow y_l,l=1...p$, we consider the corresponding tensorial structures $r^{(m)}_k$. As part of vertex data, we specify a rule to contract the indices in such a way as to obtain a form of certain rank on $w$ the manifold $M^k$.

4) Choose an immersed manifold $N \subset M^k$ of dimension $w$ and pull back the form from 3) to this manifold. Integrate this form on the fundamental class of $N$.

This construction depends on the choice of immersions $N \subset M^k$, and we obtain a much richer theory.

\section{Conclusion}

We investigated the analogies of Bethe-Salpeter equations on arbitrary Riemannian manifolds. We constructed a set of models which are very natural from geometric point of view and can be constructed on arbitrary Riemannian manifold. There is a natural hierarchy of multiparticle wave functions that can be defined on the manifold. The hierarchy depends on the choice of certain data that involves combinatorics of interaction of particles in our model. This data is discrete. We also define a more general class of models in which forms of intermediate rank on the interaction manifold of the vertices are allowed. This class of models depend on choice of immersed manifolds for each interaction manifold.

\bibliographystyle{amsplain}
\bibliography{bibliogr}

\end{document}